\documentclass[manuscript,screen]{acmart}
\AtBeginDocument{%
  }

\usepackage{pifont}
\newcommand\red[1]{\textcolor[rgb]{1,0,0}{#1}}
\newcommand\green[1]{\textcolor[rgb]{0,1,0}{#1}}
\newcommand{\CodeIn}[1]{{\small\texttt{#1}}}
\usepackage[most]{tcolorbox}
\usepackage{multirow}

\begin{document}

\title{Understanding Automated Web GUI Testing: An Empirical Study Across Exploration Strategies and State Abstractions}

\author{Chenxu Liu}
\email{chenxuliu@foxmail.com}
\affiliation{%
  \institution{SCS, Peking University; Key Lab of HCST (PKU), MOE}
  \city{Beijing}
  \country{China}
}

\author{Wei Yang}
\email{wei.yang@utdallas.edu}
\affiliation{%
  \institution{University of Texas at Dallas}
  \city{Richardson}
  \country{USA}
}
\author{Ying Zhang}
\email{zhang.ying@pku.edu.cn}
\authornotemark[1]
\affiliation{%
  \institution{Key Lab of HCST (PKU) \& National Engineering Research Center of Software Engineering, MOE; Peking University}
  \city{Beijing}
  \country{China}
}

\author{Tao Xie}
\email{taoxie@pku.edu.cn}
\authornote{Corresponding authors.}
\affiliation{%
  \institution{Key Lab of HCST (PKU), MOE; SCS, Peking University}
  \city{Beijing}
  \country{China}
}



\begin{abstract}
Automated web GUI testing (AWGT) relies on effective exploration strategies to exercise web applications through GUI actions to maximize code coverage. In recent years, a wide range of approaches have been proposed, including traditional model-based and reinforcement learning (RL)–based approaches, as well as emerging large language model (LLM)–based approaches. Meanwhile, state abstraction has long been recognized as a critical factor in guiding the exploration of AWGT approaches, detecting pages with the same functionalities from the testing perspective to avoid repeated testing. However, the impact of state abstraction across different exploration strategies remains underexplored. As a result, practitioners and researchers lack a clear understanding of how exploration strategies and state abstractions jointly affect testing effectiveness.

To bridge this gap, in this article, we present an empirical study of automated web GUI testing that analyzes exploration strategies and state abstractions from the perspectives of code coverage and failure revelation, offering actionable insights for selecting and designing effective AWGT approaches. We first compare representative model-based, RL–based, and LLM-based AWGT approaches in terms of exploration effectiveness. We then investigate how six different state abstractions influence the effectiveness of model-based and RL–based approaches. Furthermore, we examine LLM-based approaches by varying history representations, which serve as a form of state abstraction for LLM-driven approaches. Finally, we analyze and compare the failures exposed by different approaches to understand their behavioral differences.

Our empirical results indicate that no single category of exploration strategy consistently excels across all evaluation dimensions. Instead, different categories exhibit complementary strengths in code coverage, state coverage, and failure discovery. State abstraction is a key factor in testing effectiveness, with strict and fine-grained abstractions favoring model-based strategies, while compact abstractions better support RL-based strategies. Furthermore, history representation substantially impacts LLM-based strategies, where concise, functionality-level context leads to the strongest effectiveness. We also find that code coverage is not strongly correlated with failure-revealing ability, underscoring the need for multi-dimensional evaluation. Overall, our findings offer practical guidance for selecting exploration strategies, designing effective state abstractions, and improving the evaluation and effectiveness of AWGT approaches.

\end{abstract}





\maketitle

\section{Introduction}
Web applications play an important role in modern society and are widely used in domains such as news dissemination, corporate websites, and e-commerce platforms. In recent years, web applications have also become a primary deployment medium for large language models, offering interactive services to the public through chat-based interfaces. As a result, ensuring the quality and reliability of web applications become critical. To reduce the high cost of manual testing, automated web GUI testing (AWGT) approaches have been widely adopted. These approaches simulate human interactions (e.g., click, input) with web applications through graphical user interfaces (GUIs), aiming to achieve high code coverage and reveal as many failures as possible within a given time budget.

\begin{figure}[t]
    \centering
    \includegraphics[width=0.6\linewidth]{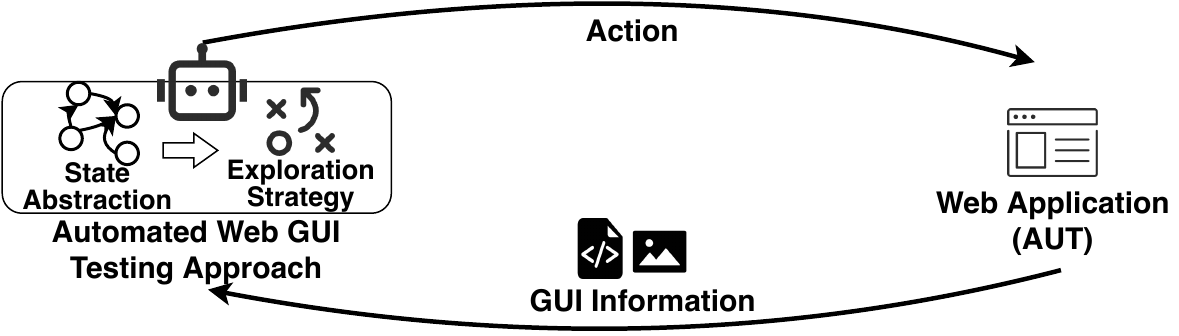}
    \caption{The workflow of a typical AWGT approach.}
    \label{fig:AWGT}
\end{figure}

A typical AWGT approach follows a cyclic workflow, as illustrated in Figure~\ref{fig:AWGT}. First, starting from an initial page of the application under test, the approach inspects the current page and collects GUI-related information, such as HTML content and screenshots. Second, the current page is incorporated into an evolving state transition graph, where the approach determines to which state the page belongs through state abstraction. 
Third, based on the adopted exploration strategy and the identified state information, the approach decides the next GUI action to execute. 
Finally, the selected action is dispatched to the application under test, leading to a new page and initiating the next exploration cycle. This process continues until the time budget is exhausted or the approach determines that the application has been sufficiently explored.

Existing AWGT approaches fall into four main categories based on their exploration strategies.
First, random-based approaches~\cite{monkey,visualRandom}. These approaches, also known as dumb monkeys~\cite{testar1}, generate GUI actions randomly without considering page semantics. While simple, these approaches often produce a large number of ineffective actions, such as clicking on non-interactive regions.
Second, model-based approaches~\cite{crawljax,testar,ATUSA,fraggen,Judge}. These approaches incrementally construct a state transition graph during testing, where each state represents pages that are functionally equivalent from the testing perspective (often referred to as near-duplicate pages~\cite{NDStudy,Judge,webembed}). Traversal algorithms, such as depth-first search (DFS), are then applied to explore the constructed graph.
Third, reinforcement learning (RL)–based approaches~\cite{WebExplor,WebQT,QExplore,WebRLED}.  These approaches also maintain a state transition graph but leverage reinforcement learning algorithms, such as Q-learning~\cite{Q-Learning} or deep Q-networks (DQN)~\cite{DQN}, to guide exploration based on rewards obtained from executed actions.
Fourth, large language model (LLM)–based approaches~\cite{AutoE2E,naviqate,VETL,temac}. These approaches exploit the strong reasoning capabilities of LLMs to generate meaningful inputs during testing, for example, to successfully complete form submissions and reach deeper application states, or even to entirely delegate action selection to the LLM.

In parallel, recent studies highlight the importance of state abstraction in guiding effective exploration, particularly for mitigating state explosion and redundant testing. Existing state abstraction techniques can be broadly grouped into two main categories.
First, threshold-based techniques~\cite{NDStudy,fraggen,WebExplor}. These techniques determine whether pages belong to the same testing state by comparing their structural or visual similarity, such as DOM text similarity or screenshot-based visual similarity, against predefined thresholds.
Second, deep learning–based techniques~\cite{webembed,Judge}. These techniques encode pages into vector embeddings using neural networks and apply supervised learning classifiers, such as support vector machines, to determine page equivalence.

Both exploration strategies and state abstractions are crucial for improving the effectiveness of AWGT approaches. An ineffective exploration strategy can cause an AWGT approach to become trapped in local exploration tarpits~\cite{vet} or fail to thoroughly exercise application functionalities, resulting in low code coverage. Similarly, an inappropriate state abstraction can lead an AWGT approach to mistakenly treat previously explored pages as new ones, causing redundant testing, or to incorrectly regard unseen pages as already tested, thereby missing testing opportunities.

Despite their importance, the relationship between exploration strategies and state abstractions remains insufficiently understood. Existing studies typically propose isolated improvements to either exploration strategies or state abstraction techniques, rather than systematically examining their combined effects on testing effectiveness. Moreover, to the best of our knowledge, the only prior empirical study on automated web GUI testing, NDStudy~\cite{NDStudy}, investigates only differences among state abstractions within a single AWGT approach, Crawljax~\cite{crawljax}. NDStudy assesses the quality of state transition graph construction by manually clustering explored pages, instead of employing objective metrics such as code coverage to evaluate testing effectiveness. Consequently, comprehensive empirical evidence is still lacking on how exploration strategies and state abstractions jointly influence the effectiveness of automated web GUI testing.

To bridge this gap, in this article, we present an empirical study of automated web GUI testing that analyzes the joint effects of exploration strategies and state abstractions from the perspectives of code coverage and failure revelation, offering actionable insights for selecting and designing effective AWGT approaches. 
Specifically, first, we compare five representative existing AWGT approaches categorized by their exploration strategies. 
Second, we integrate six different state abstraction techniques with representative model-based and RL–based AWGT approaches to examine their impact. 
Third, we incorporate four different history representations, which serve as state abstractions, into a representative LLM-based AWGT approach and study their effects. 
Finally, we compare the abilities of different AWGT approaches to expose unique failures, providing an alternative perspective on testing effectiveness.

Our results show that no single category of AWGT approaches consistently outperforms others across all evaluation dimensions. Instead, different categories demonstrate complementary strengths in code coverage, state coverage, and failure discovery. We find that state abstraction plays a crucial role in testing effectiveness, and that different abstraction strategies are better suited to different exploration paradigms. Strict and fine-grained abstractions benefit model-based approaches, whereas compact abstractions enhance the stability of RL-based approaches. Moreover, history representation substantially influences LLM-based approaches, with concise, functionality-level context yielding the best effectiveness. We also observe that code coverage does not strongly correlate with failure-revealing ability, highlighting the importance of evaluating both aspects. Overall, our findings provide actionable guidance for selecting exploration strategies, designing state abstractions, and evaluating AWGT approaches to improve testing effectiveness.

In summary, this article makes the following main contributions: 
\begin{itemize}
    \item We conduct the first study of automated web GUI testing, aiming to examine the joint effects of exploration strategies and state abstractions on testing effectiveness from the perspectives of code coverage and failure revelation. Our evaluation framework, raw data, and results are publicly available~\cite{OpenSourceRepo}.

    \item We provide a comprehensive comparison of five representative AWGT approaches across exploration strategies, including model-based, RL–based, and LLM–based approaches.
    
    \item We analyze the impact of state abstractions on traditional AWGT approaches by integrating six representative state abstraction techniques with two AWGT approaches, identifying the most suitable state abstraction technique for different exploration strategies.
    
    \item We investigate state abstraction in LLM-based AWGT approaches by studying how different history representations influence the effectiveness of LLM-based testing.
    
    \item We compare the failure-revealing capabilities of different AWGT approaches with different state abstraction techniques, evaluating the testing effectiveness from a different aspect.
    
\end{itemize}

\section{Background}
This section provides key background information necessary for understanding our study.

\subsection{Exploration Strategy}
An exploration strategy defines how an AWGT approach decides which GUI action to execute at each step, and is commonly regarded as the core component of an AWGT approach. Existing AWGT approaches are typically categorized into four main categories based on their exploration strategies.

\textbf{Random-based} approaches were first introduced by Google Monkey~\cite{monkey} on the Android platform. These approaches randomly select an action from the set of available actions according to predefined probabilities, and dispatch the selected action to a randomly chosen location on the GUI. The underlying intuition is inspired by the infinite monkey theorem, which suggests that given unlimited time, a monkey randomly typing on a typewriter could eventually produce the complete works of Shakespeare. In the context of GUI testing, this intuition translates to the idea that randomly dispatching GUI actions could eventually achieve full coverage of the application under test.

However, in practical testing scenarios, the available time budget is typically limited. As a result, random-based strategies often generate a large number of ineffective actions, such as clicking on non-interactive or blank regions of a page, leading to low code coverage. To mitigate this limitation, subsequent work has proposed restricting action dispatching to valid interactive components~\cite{WCTester,WCTester1} or leveraging visual analysis techniques to identify actionable GUI elements~\cite{visualRandom}.

\textbf{Model-based} approaches, with Crawljax~\cite{crawljax} as a representative example, incrementally construct a state transition graph during testing to serve as a model of the application under test and guide exploration. Taking Crawljax as an example, after observing a page of the application under test, Crawljax assigns the page to a state in the state transition graph and checks whether there exist interactive components on the page that have not yet been exercised. If such components exist, Crawljax either sequentially or randomly selects one component and interacts with it. Otherwise, Crawljax searches the state transition graph for a state that has not been fully explored and attempts to navigate to that state to continue testing. The testing process terminates when Crawljax believes that all states in the graph have been fully explored.

The use of a state transition graph brings two main advantages. First, pages that exhibit the same functionality from a testing perspective are abstracted into the same state, avoiding redundant testing and unnecessary time overhead. Second, the state transition graph provides a structured representation of the application under test, enabling the application of systematic traversal algorithms, such as depth-first search (DFS), to guide exploration, rather than relying on purely random testing.

\textbf{RL-based} approaches are similar to model-based approaches in that they also rely on constructing a state transition graph to guide testing. The key difference lies in how actions are selected. Instead of following predefined traversal rules, RL-based approaches determine each action using reinforcement learning algorithms, such as Q-learning~\cite{Q-Learning}.

Taking WebExplor~\cite{WebExplor} as an example, when a page is visited for the first time, WebExplor assigns an equal initial probability to each interactable component on the page. If interacting with a component leads to the discovery of a new state, WebExplor assigns a positive reward to the corresponding action and increases the probability of selecting that component in future interactions. Conversely, if a component is repeatedly interacted with but fails to reveal new states, its selection probability is gradually decayed to encourage exploration of other components.

Although existing RL-based approaches~\cite{WebQT,QExplore} adopt different reward functions, they generally share a common design principle that gives positive rewards to actions that lead to the discovery of new states. More recently, some studies have explored the use of deep Q-networks (DQNs) to guide exploration~\cite{WebRLED}. Unlike approaches that explicitly assign pages to discrete states, these DQN-equipped approaches encode pages into vector embeddings using deep neural networks and implicitly leverage state information as input for action selection.

\textbf{LLM-based} approaches leverage the strong reasoning capabilities and application-understanding abilities of large language models to perform testing. Some existing studies do not rely entirely on LLMs but integrate LLMs into model-based or RL-based approaches to assist with generating meaningful and valid inputs. This is particularly useful for overcoming barriers related to form submission or searching, thereby enabling deeper exploration of the application and achieving higher code coverage~\cite{VETL,QTypist,InputBlaster} than using only the baseline approaches. In addition, LLMs have been employed to help baseline approaches escape exploration tarpits~\cite{vet} when exploration strategies of the baseline approaches become ineffective~\cite{llmdroid}.

With the rapid improvement of LLM capabilities and the continuous reduction in inference costs, an increasing number of studies have focused on fully LLM-driven exploration, where the LLM is responsible for deciding each action step during testing. Motivated by the observation that existing model-based and RL-based approaches often fail to concentrate on exercising a specific functionality and frequently interrupting execution and switching to unrelated pages, GPTDroid~\cite{GPTDroid} models the GUI testing process as an iterative questioning-and-answering loop. In this loop, the LLM iteratively observes the current page and executes actions to cover the functionalities of the application under test. Other approaches further decompose the testing process by first summarizing application functionalities and subsequently instructing the LLM to focus on executing one functionality at a time~\cite{AutoE2E,naviqate,DroidAgent,mobileGPT,temac}.

\subsection{State Abstraction}
\label{sec:state abstraction}
Except for random-based approaches, the other three categories of AWGT approaches rely heavily on state abstraction to effectively conduct testing. An ineffective state abstraction can substantially impair the testing effectiveness of AWGT approaches. The goal of state abstraction is to group pages that exhibit the same functionalities from the testing perspective into the same abstract state.
If pages with the same functionalities are incorrectly treated as different, exploration strategies can be misled into repeatedly testing equivalent pages, resulting in redundant exploration and wasting testing time, consequently reducing code coverage. Conversely, if pages with different functionalities are incorrectly merged into the same state, exploration strategies can overlook potential testing opportunities, leading to insufficient exploration and reducing code coverage.
State abstraction techniques are typically categorized into two main categories.

\textbf{Similarity-based} techniques determine whether two pages belong to the same abstract state by computing their similarity based on either HTML textual content or screenshot visual appearance, and comparing the resulting similarity score against a predefined threshold. Representative HTML-based techniques include RTED~\cite{RTED}, Levenshtein ratio~\cite{Levenshtein}, and SimHash~\cite{SimHash}, while visual–based techniques include PDiff~\cite{PDiff}, PHash~\cite{PHash}, and SIFT~\cite{SIFT}.

Given a newly observed page, these techniques compute its similarity with the representative pages of each existing state. If at least one of the similarity scores exceeds the predefined threshold, the page is assigned to the state with the highest similarity. Otherwise, if none of the similarity scores exceed the threshold, a new state is created for the page.

In recent years, several variants of threshold-based state abstraction techniques have been proposed. FragGen~\cite{fraggen} represents a page by decomposing its screenshot into a set of component-level fragments based on page layout. During state comparison, FragGen performs a set-based comparison between the fragment sets of two pages to determine whether they should be assigned to the same state. This process can be viewed as a similarity comparison with an implicit threshold of zero. WebExplor~\cite{WebExplor} adopts a hybrid technique that combines heuristic rules with similarity comparison. Specifically, it first applies rule-based heuristics to treat any page associated with a newly encountered URL as a new state. It then performs a similarity comparison based on Gestalt pattern matching~\cite{Gestalt} of HTML tags to further refine state assignment.

\textbf{Deep learning–based} techniques leverage unsupervised deep neural networks to encode web pages into vector embeddings, and subsequently train a supervised machine learning classifier that takes the embeddings of two pages as input to determine whether the two pages should be assigned to the same abstract state. By performing decision making in a high-dimensional feature space, these techniques can incorporate rich semantic information of web pages, enabling effective state abstraction~\cite{TK,webembed,Judge}.

\section{Empirical Study Design}
This empirical study is structured to answer the following four research questions:
\begin{itemize}
    \item \textbf{RQ1 (Overall Effectiveness)}: To what extent and how do different exploration strategies for automated web GUI testing compare in terms of test effectiveness?
    \item \textbf{RQ2 (Effect of State Abstractions)}: To what extent and how do different state abstractions influence the test effectiveness of model-based and RL–based exploration strategies?
    \item \textbf{RQ3 (Effect of History Representations)}: To what extent and how do different history representations, as a form of state abstraction, affect the test effectiveness of LLM-based exploration strategies?
    \item \textbf{RQ4 (Failure-Revealing Capability)}: To what extent and how do different exploration strategies and state abstraction techniques differ in their failure-revealing ability?
\end{itemize}

\subsection{AWGT Approaches for Comparison}
In this empirical study, we include representative AWGT approaches with five different exploration strategies for comparison. We excluded random-based approaches since they do not benefit from state abstraction techniques. The AWGT approaches are listed as follows. 

\begin{itemize}
    \item \textbf{Crawljax}~\cite{crawljax}. A model-based AWGT approach that has been adopted as a baseline in many prior AWGT approaches~\cite{FeedEx,WebExplor,QExplore,temac}, and has been widely used as a baseline approach for the studies of state abstraction techniques~\cite{fraggen,webembed,Judge}.
    \item  \textbf{FragGen}~\cite{fraggen}. A model-based AWGT approach that is developed based on Crawljax, but equipped with a fragment-based state abstraction technique.
    \item \textbf{WebExplor}~\cite{WebExplor}. An RL–based AWGT approach that adopts Q-learning~\cite{Q-Learning} as its exploration strategy. WebExplor has also been widely adopted as a baseline in existing work~\cite{QExplore,WebQT,WebRLED}.
    \item \textbf{WebRLED}~\cite{WebRLED}. An RL-based AWGT approach that uses DQN~\cite{DQN} to guide exploration. 
    \item \textbf{GPTWeb}~\cite{GPTDroid}. An LLM-based AWGT approach that leverages an LLM to decide each action to dispatch during testing. GPTWeb is developed by us based on GPTDroid~\cite{GPTDroid}. We made modifications and improvements to migrate GPTDroid to the web scenario. We do not adopt an existing LLM-based AWGT approach because the existing AWGT approaches are either designed for only generating textual inputs~\cite{VETL}, or focus on inferring application features rather than improving code coverage~\cite{AutoE2E,naviqate}.
\end{itemize}

For Crawljax, FragGen, and WebRLED, we use their publicly available implementations. For WebExplor, we obtain the source code by contacting the authors. For all baseline approaches, we configure them using the recommended settings provided by the respective authors. Since Crawljax (and FragGen, which is built on top of Crawljax) tends to terminate before exhausting the given time budget, we modify Crawljax to automatically restart its exploration until the time budget is fully utilized, following the experimental settings adopted in Judge~\cite{Judge}, for a fair comparison.

For GPTWeb, since it is originally designed for the Android platform (GPTDroid) and cannot be directly applied to web applications, we adapt it to the web scenario through a series of modifications and enhancements. To facilitate reproducibility and support future research, we release GPTWeb in our open-source repository~\cite{OpenSourceRepo}.
Specifically, we make the following four modifications.
First, as the prompts used in GPTDroid are tailored to the Android platform, we revise them to suit the web setting. We retain the overall prompt framework of GPTDroid, including the initialization and testing prompts, operation and questioning patterns, and the functionality summarization workflow. Meanwhile, we replace Android-specific contents with their web counterparts (e.g., replacing Android activity names with URLs), and incorporate additional contextual information from the prompt templates used in UGround~\cite{uground} to enrich the model’s understanding of web applications.
Second, we adapt the action space of GPTDroid to the web scenario by aligning it with the WebArena framework~\cite{webarena}. Accordingly, we replace the underlying action dispatcher from Android Debug Bridge (ADB) to Playwright to support web-based GUI interactions.
Third, we equip GPTWeb with a multi-modal and planner–actor decoupling design inspired by SeeAct~\cite{SeeAct} and UGround~\cite{uground}. Instead of passing the entire HTML of a web page to the LLM and requiring it to select an action via XPath, we first provide the page screenshot and testing instructions to an LLM acting as a planner (GPT-4o\footnote{https://platform.openai.com/docs/models/gpt-4o} in our implementation), which outputs a natural language description of the next GUI component to interact with. This description is then passed to another LLM acting as an actor (UI-TARS-1.5-7B~\cite{ui-tars-15-seed} in our implementation), which grounds the target component to a concrete pixel-level location to dispatch the corresponding action.
As web pages are typically much more complex than mobile pages~\cite{VETL}, replacing HTML/XML-based state representations with screenshots substantially mitigates the long-context issue~\cite{LostInTheMiddle}, thereby improving testing effectiveness.
Fourth, we integrate GPTWeb with four different history representations, which serve as forms of state abstraction, to study the impact of state abstraction on the effectiveness of LLM-based testing. Further details of the history representations are provided in Section~\ref{sec:setup-state abstraction}.

In our study, we exclude TESTAR~\cite{testar} and GUITAR~\cite{guitar}, as Crawljax~\cite{crawljax} has been widely studied in recent work and supports integration with multiple state abstraction techniques, making it sufficient to represent model-based AWGT approaches in our evaluation. We exclude QExplore~\cite{QExplore}, as limitations in its publicly available infrastructure for dispatching actions prevent it from performing effective testing in our experimental setting. We exclude AutoE2E~\cite{AutoE2E} and NaviQAte~\cite{naviqate} since they do not focus on improving code coverage. We also exclude WebQT~\cite{WebQT} and VETL~\cite{VETL} because they are not open-sourced.

According to existing work~\cite{NDStudy,WebExplor,WebQT,WebRLED,Judge}, AWGT approaches exhibit randomness during testing. To mitigate this randomness, we repeat each experiment five times and report the average results for our RQs, adhering to existing work~\cite{WebQT,WebRLED}. For all experiments, we set a 30-minute time budget, which is demonstrated to be sufficient for existing AWGT approaches to reach coverage stagnation~\cite{WebExplor,WebQT,Judge}. To ensure a fair comparison, we set a fixed interval of 1,000 ms between consecutive actions and apply the same login scripts across all approaches.

\subsection{State Abstraction Techniques for Comparison}
\label{sec:setup-state abstraction}
To study the impact of state abstraction on AWGT approaches, we include six different state abstraction techniques for the model-based and RL-based AWGT approaches, and four different history representation strategies for LLM-based AWGT approaches.

The state abstraction techniques are listed as follows.
\begin{itemize}
    \item \textbf{String comparison (StringCmp)}~\cite{crawljax}.  
    The default state abstraction technique adopted by Crawljax~\cite{crawljax}, determining page equivalence based on string-level comparisons of HTML content.
    \item \textbf{Gestalt pattern matching (Gestalt)}~\cite{WebExplor}.  
    The default state abstraction technique adopted by WebExplor~\cite{WebExplor}, combining heuristic rules with Gestalt-based similarity matching.
    \item \textbf{Robust tree edit distance (RTED)}~\cite{RTED,NDStudy}.  
    A threshold-based state abstraction technique that is reported by NDStudy~\cite{NDStudy} as the most effective HTML-based technique for state abstraction.
    \item \textbf{Perceptual difference (PDiff)}~\cite{PDiff,NDStudy}.  
    A threshold-based state abstraction technique that is reported by NDStudy~\cite{NDStudy} as the most effective screenshot-based technique for state abstraction.
    \item \textbf{WebEmbed}~\cite{webembed}.  
    A deep learning–based state abstraction technique that encodes web pages into vector embeddings. WebEmbed is the first deep learning-based technique.
    \item \textbf{Judge}~\cite{Judge}.
    A deep learning–based state abstraction technique that is shown to be the most effective in guiding Crawljax for automated web GUI testing.
\end{itemize}

We integrate the selected state abstraction techniques with Crawljax and WebExplor, which represent model-based and RL-based AWGT approaches, respectively. We choose Crawljax because a large body of prior work~\cite{NDStudy,fraggen,webembed,Judge} on state abstraction for AWGT are built upon Crawljax. We select WebExplor because it explicitly constructs and leverages a state transition graph during exploration, in contrast to WebRLED, which relies on vector embeddings of pages to implicitly encode state information without explicitly modeling state transitions.

Since some state abstraction techniques are implemented in Java while others are implemented in Python, to ensure implementation consistency and fair comparison, we encapsulate all state abstraction techniques that are not natively supported by the AWGT approaches as independent Flask services, and invoke them via network requests during testing.
Specifically, Crawljax natively supports StringCmp, RTED, and PDiff, and invokes the remaining state abstraction techniques via network requests. WebExplor natively supports StringCmp and Gestalt, and similarly relies on network requests to invoke the remaining state abstraction techniques.

For a fair comparison, we adopt the recommended configurations~\cite{NDStudy,WebExplor,fraggen,webembed,Judge} and open-source implementations~\cite{WebExplor,fraggen,webembed,Judge} of the state abstraction techniques. Specifically, we adopt the default classification threshold of 0.8 for Gestalt. We choose the medium similarity value of the ``near-duplicate'' class in a labeled dataset provided by NDStudy~\cite{NDStudy} as the classification threshold of RTED and PDiff. We adopt the pre-trained models and classifiers of WebEmbed and Judge.

The history representation strategies are listed as follows.
\begin{itemize}
    \item \textbf{No history}~\cite{no_history}.  
    The most basic strategy, in which no historical information is retained, and the LLM makes each action decision based on only the current page.
    \item \textbf{Action history}~\cite{SeeAct,uground}.  
    A strategy that preserves the sequence of previously dispatched actions as historical context.
    \item \textbf{State history (action history with states)}~\cite{action_state_history}.  
    A strategy that retains both action information and a summary of the resulting state after each action, interleaving actions with state descriptions in the history.
    \item \textbf{Functionality history}~\cite{GPTDroid}.  
    The default history representation strategy used in GPTDroid~\cite{GPTDroid}, which requires the LLM to summarize the functionalities being exercised after each action and maintain these summaries as historical context.
\end{itemize}

We select these representative history representation strategies based on prior work on GUI agents, and integrate them into GPTWeb to investigate the impact of state abstraction on the effectiveness of LLM-based AWGT approaches.
Specifically, for action information, we record each dispatched action using the triplet \CodeIn{[Action Type, Target Element, Value]}. For state information, we employ GPT-4o, taking page screenshots as input and prompting it to produce a natural language description of the current page, following previous work~\cite{temac}. For functional information, we follow the original implementation and prompts of GPTDroid to require the model to summarize the exercised functionalities as historical context. The templates and instantiations for each type of history representation are shown in Table~\ref{tab:template and example}.

\begin{table*}[t]
\caption{Templates and instantiations for each type of history representation.}
\begin{center}
\resizebox{\textwidth}{!}{
\begin{tabular}{@{}p{1.5cm}|p{4cm}|p{13cm}@{}}
\toprule
\textbf{Type} & \textbf{Templates} & \textbf{Instantiations} \\
\midrule
Action & \textit{\textless Action type\textgreater \ + \textless Component\textgreater \ + \textless Value\textgreater} & \small Latest 1 action taken is Type on element with description: the NAME input field in the 'Add Menu' dialog using value: New Menu. \\
\midrule
Action with states & \textit{\textless URL\textgreater \ + \textless Description\textgreater \ + \textless Action\textgreater \ + \textless Transition Action\textgreater} & \small Latest 1 step tested the http://localhost:4004/session/rJeWapm0gx page. The page is about The webpage is part of a retrospective tool titled "Retrospected," designed for Agile meetings, with sections for users to input feedback on what went well, what could be improved, and any brilliant ideas to share; it is currently in a blank state with no entries made, and the user logged in is identified as "TestUser.". The action taken is type on element with description: the text input field under 'A brilliant idea to share?' using value: Implementing a weekly team-building exercise. The following executive command achieve the page transition: Action: click on element with description: the pink CREATE A NEW SESSION button using value: . \\
\midrule
Functional & \textit{\textless Functionality\textgreater \ + \textless Visits\textgreater \ + \textless Status\textgreater \newline \textless URL\textgreater \ +  \textless Action\textgreater \ + \textless Transition Action\textgreater} & \small Function 1: Add wallet. Visits: 1. Status: In Progress \newline Lastest 1 step tested the http://localhost:4000/wallets page. The action taken is click on element with description: the 'Add' button below the Test Wallet using value: . The following executive command achieve the page transition: Action: click on element with description: the 'Next' button in the tour step using value: . \\ 
\bottomrule
\end{tabular}
}
\label{tab:template and example}
\end{center}
\end{table*}

\subsection{Application Subjects}
We adopt six open-source web applications that have been widely examined in prior studies~\cite{fraggen,WebExplor,WebQT,webembed,WebRLED} as subjects for our evaluation. These applications were originally introduced by DIG~\cite{DIG}. They were collected in 2019 and instrumented using the front-end code instrumentation tool NYC~\cite{NYC}, with branch coverage used as the coverage metric. Detailed information about the selected applications is provided in Table~\ref{tab:webapps_info}, where LOC denotes lines of code.

\begin{table}[t]
    \centering
    \caption{Detailed information of our application subjects.}
    \resizebox{0.6\linewidth}{!}{
    \begin{tabular}{@{}c|cccc@{}}
        \toprule
        \textbf{Name}   & \textbf{Version} & \textbf{Client LOC} & \textbf{Server LOC} & \textbf{Domain}  \\
        \midrule
        Dimeshift~\cite{Dimeshift} & 2018 & 5,140 (JS) & 3,298 (JS) & Finance  \\

        Pagekit~\cite{Pagekit} & 1.0.15 & 4,214 (JS) & 13,856 (PHP) & Publishing  \\

        Petclinic~\cite{Petclinic}  & 2018   & 2,939 (JS) & 842 (Java) & Healthcare \\
        
        Phoenix~\cite{Phoenix} & 1.1 & 2,289 (JS) & 1,135 (Elixir) & Management \\
        
        Retroboard~\cite{Retroboard} & 2019 & 2,144 (JS) & 278 (JS) & Collaboration \\ 
        
        Splittypie~\cite{Splittypie} & 2018 & 2,710 (JS) & 829 (JS) & Finance  \\
        \bottomrule
    \end{tabular}
    
    }
    \label{tab:webapps_info}
\end{table}

\subsection{Evaluation Environment}
We conduct our evaluation on an Ubuntu 20.04 PC equipped with an Intel i5-12400F CPU and 32 GB of RAM. The Chromium driver version is 112.0.5615.165.

\section{RQ1 Overall Effectiveness}
\begin{figure}[]
    \centering
    \includegraphics[width=0.87\linewidth]{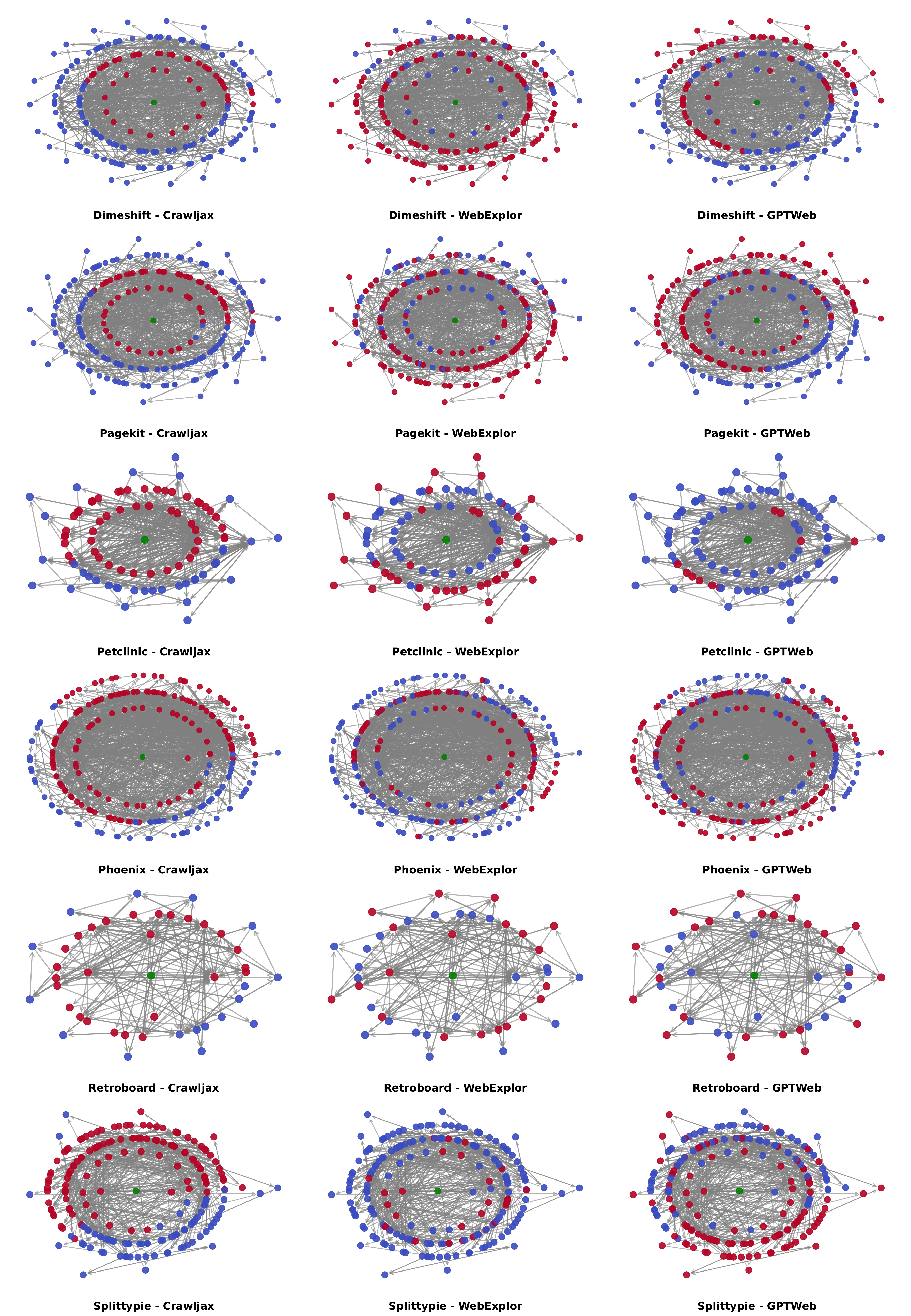}
    \caption{The state transition graph of each application, colored by the coverage status of each AWGT approach.}
    \label{fig:RQ1Graph}
\end{figure}

In RQ1, we compare the code coverage achieved by AWGT approaches representing different categories and further perform a detailed state-level analysis to understand how and why their testing effectiveness differs. The code coverage achieved by each approach is shown in Table~\ref{tab:RQ1}, and the state-level analysis is shown in Figure~\ref{fig:RQ1Graph}.

\subsection{Analysis of Code Coverage}

As shown in Table~\ref{tab:RQ1}, RL-based approaches achieve the highest coverage most frequently and obtain the highest average coverage overall. However, no single category consistently outperforms the others across all applications, as testing effectiveness is jointly influenced by application-specific page characteristics and the design of the exploration strategy.

\begin{table}[t]
\centering
\caption{Code coverage of approaches from different categories.}
\resizebox{0.7\linewidth}{!}{
\begin{tabular}{@{}c|cc|cc|c@{}}
\toprule
\textbf{}            & \multicolumn{2}{c|}{\textbf{Model-based}}                                                                           & \multicolumn{2}{c|}{\textbf{RL-based}}      & \textbf{LLM-based}                                                                    \\ \midrule
\textbf{Application} & \textbf{Crawljax} & \textbf{FragGen} & \textbf{WebExplor} & \textbf{WebRLED} & \textbf{GPTWeb} \\ \midrule
Dimeshift            & 34.57\%           & 37.96\%          & \textbf{53.50\%}   & 53.09\%          & 36.72\%         \\
Pagekit              & 33.41\%           & \textbf{37.73\%} & 24.05\%            & 24.00\%          & 32.08\%         \\
Petclinic            & 66.00\%           & 61.00\%          & \textbf{85.00\%}   & \textbf{85.00\%}          & 54.00\%         \\
Phoenix              & 67.11\%           & 66.84\%          & \textbf{69.74\%}   & 69.21\%          & 69.47\%         \\
Retroboard           & 53.80\%           & 55.56\%          & 59.77\%            & \textbf{69.24\%}          & 62.81\%         \\
Splittypie           & 39.83\%           & 43.50\%          & 22.65\%            & \textbf{45.04\%}          & 41.28\%         \\ \midrule
Average           & 49.12\%           & 50.43\%          & 52.45\%            & \textbf{57.60\%}          & 49.39\%         \\ \bottomrule
\end{tabular}}
\label{tab:RQ1}
\end{table}

Surprisingly, despite leveraging the powerful reasoning capabilities of LLMs, the LLM-based AWGT approach GPTWeb does not achieve high code coverage. We attribute this result to two primary factors.
First, LLM-based approaches suffer from low execution efficiency. Each decision step in GPTWeb requires invoking a large language model, which introduces substantial computational overhead. As a result, GPTWeb typically executes fewer than 300 action steps within a 30-minute time budget, being three to four times fewer than model-based and RL-based approaches. This inefficiency greatly reduces GPTWeb’s opportunity for trial-and-error exploration and slows coverage growth.
Second, the capability of the backbone LLM remains insufficient. We observe that the LLM frequently repeats failed or invalid actions, wastes time generating improperly formatted input text, or incorrectly grounds the locations of GUI elements. These issues are particularly severe in Dimeshift and Petclinic, leading to substantially lower code coverage. We expect these limitations to diminish as LLM capabilities continue to improve.

RL-based AWGT approaches demonstrate their advantage in employing more effective exploration strategies than model-based approaches, while avoiding the substantial efficiency overhead incurred by LLM-based approaches. However, WebExplor and WebRLED perform poorly on Pagekit due to two key factors.
First, RL-based approaches tend to over-prioritize components encountered early in the exploration process. Pagekit features a landing page that yields limited coverage gains, whereas deeper functional and settings pages offer substantially higher coverage potential. Early positive rewards obtained from interactions on the landing page bias the Q-values, causing the approach to over-explore the landing page for an extended period.
Second, RL-based approaches struggle with complex form interactions. The settings page of Pagekit contains numerous single-choice and multi-choice components, where maximizing coverage requires systematically exploring diverse input combinations. However, probabilistic action selection based on Q-tables discourages exhaustive traversal of such combinations, thereby limiting coverage growth.

It is worth noting that WebExplor and WebRLED exhibit substantial differences in testing effectiveness on Splittypie due to their distinct reward update strategies. WebExplor maintains a separate Q-table for each state and updates rewards after every action step, whereas WebRLED employs a random warm-up phase and updates its DQN only after completing each epoch (at least 21 steps). This batch-style learning strategy allows WebRLED to capture broader coverage opportunities in each update cycle and reduces the risk of premature convergence to local exploration loops.
Nevertheless, for more complex applications such as Pagekit, even DQN-based strategies demonstrate limited effectiveness, suggesting that scalability remains a key challenge for RL-based AWGT approaches.

Model-based AWGT approaches follow fixed traversal strategies (e.g., DFS), making them less susceptible to the reward bias that affects RL-based approaches. This characteristic enables them to achieve notably higher coverage on Pagekit and Splittypie. However, due to their reliance on traversal-style exploration, model-based approaches often struggle with complex functionalities that require long sequences of interdependent actions, thereby imposing an upper bound on the achievable coverage.

\textbf{Answer 1.1}: RL-based approaches present the best testing effectiveness among the three categories. Although LLM-based approaches have the potential to become the most general-purpose solution as LLM capabilities continue to advance, they do not yet demonstrate clear superiority in testing effectiveness at present. In practice, testers can combine approaches from different categories to exploit their complementary strengths. 

\subsection{Analysis of State Transition Graph}

To investigate how existing AWGT approaches cover applications at the state level, we construct a state transition graph for each subject application and visualize the state coverage achieved by different approaches, as shown in Figure~\ref{fig:RQ1Graph}.
Specifically, we collect the HTML pages recorded at each execution step from five independent runs of Crawljax, WebExplor, and GPTWeb on each application (15 runs in total). We then construct a state transition graph for each application using Judge~\cite{Judge}, which has been demonstrated to be the state-of-the-art state abstraction technique.
For visualization, we place the initial state (State 0) at the center of the graph and color it green. The remaining states are arranged in concentric circles according to their shortest-path distance from the initial state, such that states farther from the entry point appear on outer rings. We highlight states visited by each AWGT approach in red and unvisited states in blue to facilitate cross-approach comparison. A state is considered visited if it is reached at least once across the five runs of an approach. Gray arrows represent transition relations between states.

As shown in Figure~\ref{fig:RQ1Graph}, AWGT approaches from different categories exhibit notable complementarity. This is particularly evident between Crawljax and WebExplor on Petclinic, as well as between Crawljax and GPTWeb on Splittypie.

Crawljax (representing model-based approaches) primarily explores states in the shallow layers near the initial state, concentrated in a relatively limited region. Although its model-driven strategy constrains its ability to execute long and complex action sequences for deep exploration, it demonstrates strong exploration breadth and superior coverage in shallow states.

WebExplor (representing RL-based approaches) achieves a more balanced trade-off between exploration breadth and depth. Even on Pagekit, where its overall coverage is relatively low, it is still able to reach states on deep levels. However, WebExplor fails to thoroughly test these states to obtain high coverage. On Splittypie, WebExplor often becomes trapped in shallow states, limiting its deep exploration.

GPTWeb (representing LLM-based approaches) can reach application states on deep levels with fewer action steps than other approaches. Except for Dimeshift and Petclinic, where LLM limitations prevent it from accessing certain states, GPTWeb successfully covers most deep-layer states (i.e., the two outermost rings) in other applications.

\begin{table}[t]
\centering
\caption{Code coverage and state coverage results of each approach on each application subjects.}
\resizebox{\linewidth}{!}{
\begin{tabular}{@{}c|cc|cc|cc@{}}
\toprule
                     & \multicolumn{2}{c|}{\textbf{Crawljax}}            & \multicolumn{2}{c|}{\textbf{WebExplor}}           & \multicolumn{2}{c}{\textbf{GPTWeb}}                       \\ \midrule
\textbf{Application} & \textbf{Code Coverage} & \textbf{State Coverage} & \textbf{Code Coverage} & \textbf{State Coverage} & \textbf{Code Coverage} & \textbf{State Coverage} \\ \midrule
Dimeshift            & 34.57\%                & 26.79\%                 & 53.50\%                & 79.43\%                 & 36.72\%                & 41.63\%                 \\
Pagekit              & 33.41\%                & 30.60\%                 & 24.05\%                & 66.38\%                 & 32.08\%                & 57.76\%                 \\
Petclinic            & 66.00\%                & 58.62\%                 & 85.00\%                & 48.28\%                 & 54.00\%                & 11.49\%                 \\
Phoenix              & 67.11\%                & 67.50\%                 & 69.74\%                & 37.50\%                 & 69.47\%                & 63.33\%                 \\
Retroboard           & 53.80\%                & 61.36\%                 & 59.77\%                & 45.45\%                 & 62.81\%                & 56.82\%                 \\
Splittypie           & 39.83\%                & 62.44\%                 & 22.65\%                & 15.23\%                 & 41.28\%                & 56.85\%                 \\ \bottomrule
\end{tabular}}
\label{tab:RQ1.2}
\end{table}

We further compute state coverage as the ratio of visited states to the total number of states and compare it with code coverage in Table~\ref{tab:RQ1.2}. We observe substantial ranking differences between code coverage and state coverage across approaches (e.g., Phoenix), and similar code coverage can correspond to markedly different state coverage (e.g., Dimeshift). This is because different states contribute unevenly to code coverage. Visiting certain key states yields large coverage gains, and actions that do not cause state changes can still increase code coverage.

It is worth noting that obtaining the ground-truth state transition graph of a web application is highly challenging. In this research question, we construct the state transition graph solely based on the states visited by three different AWGT approaches. Nevertheless, this approach is sufficient to enable a meaningful comparison of the similarities and differences among the AWGT approaches.

\textbf{Answer 1.2}: Crawljax mainly covers shallow states, WebExplor balances exploration breadth and depth, and GPTWeb tends to reach deeper states within fewer steps, showing complementarity among approaches. We find that state coverage does not directly correspond to code coverage. We recommend future work to analyze coverage through state transition graphs when compared with baseline approaches, and use the state-cover information to guide the design of AWGT approaches.

\section{RQ2 Effect of State Abstractions}
In RQ2, we investigate the impact of different state abstraction techniques on AWGT approaches from the perspective of code coverage. We integrate six representative state abstraction techniques with Crawljax and WebExplor, and report the results in Table~\ref{tab:RQ2}. For each application, we highlight in bold the highest coverage achieved by Crawljax and WebExplor, respectively.

As shown in Table~\ref{tab:RQ2}, Crawljax achieves its best effectiveness when integrated with Judge, whereas WebExplor most frequently achieves the highest coverage when using its default state abstraction technique, Gestalt, and attains the highest average code coverage when combined with WebEmbed. Although each of Crawljax and WebExplor has a state abstraction technique that yields their respective optimal effectiveness, no single state abstraction technique consistently maximizes average code coverage across both AWGT approaches. This finding indicates a strong dependency between state abstraction techniques and exploration strategies, highlighting the importance of their effective coordination.

State abstraction has a substantial impact on code coverage. For example, Crawljax exhibits a 14.79\% difference in average coverage when using PDiff compared to Judge. However, state abstraction primarily influences how effectively an AWGT approach realizes its testing potential, rather than determining its inherent upper bound. For instance, Crawljax consistently achieves lower coverage than WebExplor on Petclinic (69.00\% versus 85.00\%) and Retroboard (55.56\% versus 60.23\%), regardless of the state abstraction technique employed.

To further investigate the relationship between state abstraction and exploration strategy, we report the number of states constructed by different state abstraction techniques using HTML pages and screenshots collected from five default runs of Crawljax and WebExplor. The results are summarized in Table~\ref{tab:RQ2.1}.

\begin{table}[t]
\centering
\caption{Code coverage of Crawljax and WebExplor, integrated with six state abstraction techniques.}
\setlength{\tabcolsep}{1pt}
\resizebox{\linewidth}{!}{
\begin{tabular}{@{}c|cccccc|cccccc@{}}
\toprule
\textbf{}            & \multicolumn{6}{c|}{\textbf{Crawljax}}                                                                           & \multicolumn{6}{c}{\textbf{WebExplor}}                                                                           \\ \midrule
\textbf{Application} & \textbf{StrCmp} & \textbf{Gestalt} & \textbf{RTED}    & \textbf{PDiff}   & \textbf{WebEmbed} & \textbf{Judge}   & \textbf{StrCmp}  & \textbf{Gestalt} & \textbf{RTED}    & \textbf{PDiff}   & \textbf{WebEmbed} & \textbf{Judge}   \\ \midrule
Dimeshift            & 34.57\%         & 29.44\% & 39.74\%          & 19.57\%          & 38.74\%           & \textbf{40.33\%} & 44.74\%          & \textbf{53.50\%} & 42.43\%          & 31.20\%          & 51.67\%           & 45.35\%          \\
Pagekit              & 33.41\%         & \textbf{42.69\%} & 41.75\% & 30.41\% & 41.83\%           & \textbf{42.69\%} & 22.67\%          & \textbf{24.05\%} & 20.34\%          & 21.44\%          & 23.63\%           & 20.63\%          \\
Petclinic            & 66.00\%         & \textbf{69.00\%}          & 61.00\% & 54.00\%          & 68.00\%           & \textbf{69.00\%} & 58.00\%          & \textbf{85.00\%} & \textbf{85.00\%} & 81.00\%          & \textbf{85.00\%}  & \textbf{85.00\%} \\
Phoenix              & 67.11\%         & 44.74\%          & 69.47\%          & 47.11\% & 34.21\%           & \textbf{72.37\%} & 67.11\%          & \textbf{69.74\%} & 66.32\%          & 59.74\%          & 69.47\%           & 69.47\%          \\
Retroboard           & 53.80\%         & 52.98\%          & \textbf{55.56\%} & \textbf{55.56\%} & \textbf{55.56\%}  & \textbf{55.56\%} & 59.42\%          & 59.77\% & \textbf{60.23\%} & \textbf{60.23\%} & \textbf{60.23\%}  & \textbf{60.23\%} \\
Splittypie           & 39.83\%         & 39.83\%          & 39.15\%          & 29.66\%          & 34.96\%           & \textbf{45.13\%} & \textbf{31.71\%} & 22.65\%          & 20.26\%          & 20.77\%          & 30.68\%           & 30.00\%          \\ \midrule
Average & 49.12\%           & 46.45\%            & 51.11\%            & 39.39\%            & 45.55\%             & \textbf{54.18\%}   & 47.28\%  & 52.45\%            & 49.10\%            & 45.73\%            & \textbf{53.45\%}             & 51.78\%            \\ \bottomrule
\end{tabular}}
\label{tab:RQ2}
\end{table}

\begin{table}[t]
\centering
\caption{The number of states constructed by each state abstraction technique, based on the HTML pages and screenshots collected by Crawljax and WebExplor under their default configurations.}
\setlength{\tabcolsep}{2pt}
\resizebox{\linewidth}{!}{
\begin{tabular}{@{}c|cccccc|cccccc@{}}
\toprule
\textbf{}            & \multicolumn{6}{c|}{\textbf{Crawljax}}                                                                           & \multicolumn{6}{c}{\textbf{WebExplor}}                                                                           \\ \midrule
\textbf{Application} & \textbf{StrCmp} & \textbf{Gestalt} & \textbf{RTED}    & \textbf{PDiff}   & \textbf{WebEmbed} & \textbf{Judge}   & \textbf{StrCmp}  & \textbf{Gestalt} & \textbf{RTED}    & \textbf{PDiff}   & \textbf{WebEmbed} & \textbf{Judge}   \\ \midrule
Dimeshift            & 316             & 17               & 104           & 189            & 25                & 56             & 1358            & 16               & 604           &  102              & 29                & 166            \\
Pagekit              & 402             & 40               & 90            &  565              & 17                & 71             & 1756            & 18               & 268           &  65              & 10                & 154            \\
Petclinic            & 970             & 11               & 58            &  62              & 7                 & 51             & 719             & 12               & 75            &  15              & 10                & 42             \\
Phoenix              & 970             & 68               & 558           &  210              & 6                 & 162            & 437             & 8                & 209           & 76               & 10                & 90             \\
Retroboard           & 3774            & 279              & 48            &  274              & 6                 & 27             & 1681            & 88               & 29            &  8              & 9                 & 20             \\
Splittypie           & 1590            & 231              & 233           &  27              & 9                 & 123            & 425             & 5                & 31            &  12              & 4                 & 30             \\\bottomrule
\end{tabular}}
\label{tab:RQ2.1}
\end{table}

The model-based approach Crawljax benefits more from effective and fine-grained state abstraction techniques that construct a larger number of states. According to the evaluation results reported in Judge~\cite{Judge}, Judge, WebEmbed, and RTED all achieve high F1 scores in state abstraction. However, when duplicate pages (i.e., pages that should belong to the same state) are treated as the positive class, Judge and RTED exhibit higher precision, whereas WebEmbed attains higher recall. This indicates that Judge and RTED adopt stricter classification criteria and tend to construct more states, while WebEmbed merges pages into fewer states.
When guiding Crawljax, coarser abstractions such as WebEmbed can lead to premature termination, as Crawljax may incorrectly conclude that the application has been fully explored at an early stage. Even with our automatic restart mechanism to exhaust the entire time budget, this limitation cannot be effectively mitigated. In contrast, state abstraction techniques that construct more states (e.g., Judge) enable Crawljax to perform more fine-grained traversal, thereby achieving higher code coverage. Notably, even highly strict abstractions such as StrCmp remain effective for Crawljax. However, a comparison among StrCmp, RTED, and Judge suggests that excessive state fragmentation can introduce redundant exploration and ultimately reduce coverage, indicating that model-based AWGT approaches benefit most from state abstraction techniques that are both strict and effective.

The RL-based approach WebExplor is more robust to variations in state abstraction effectiveness and exhibits smaller coverage fluctuations than Crawljax when switching among different state abstraction techniques. Unlike Crawljax, WebExplor performs better with coarse state abstraction techniques that construct a few states, such as Gestalt and WebEmbed, because RL-based exploration strategies are better suited to operating over a compact state space.
During exploration, WebExplor maintains a separate Q-table for each state, with initially uniform interaction probabilities over UI components. When the number of states becomes excessively large, WebExplor degenerates toward nearly random exploration or overfits to a small set of early positive rewards, leading to high interaction probabilities for certain components and thereby hindering effective exploration. By maintaining a small state space, WebExplor can update Q-values effectively and achieve a favorable exploration–exploitation tradeoff~\cite{sutton1998reinforcement}.
Moreover, many applications contain shared UI elements (e.g., top navigation bars or sidebars) that appear across multiple pages. Constructing a few states enables WebExplor to share learned Q-values across structurally similar pages, reducing redundant interactions with repeated components.
Nevertheless, for Pagekit and Splittypie, varying the state abstraction strategy does not mitigate the RL bias toward components encountered early in exploration, as discussed in RQ1. Consequently, WebExplor remains less effective than Crawljax on these applications, regardless of the state abstraction techniques employed.

\textbf{Answer 2}: No single state abstraction technique consistently maximizes coverage across AWGT approaches. Instead, state abstraction primarily influences how effectively an approach exploits its testing potential rather than its coverage upper bound.
Model-based AWGT approaches benefit most from strict and effective abstractions that enable fine-grained state differentiation, whereas RL-based AWGT approaches benefit from compact abstractions that support stable value updates and Q-value sharing across structurally similar pages.

\section{RQ3 Effect of History Representations}
In RQ3, we model state abstraction using four history representations and investigate their impact on the code coverage of GPTWeb. For action history, state history, and functionality history, we further examine how different history lengths (i.e., the most recent 5 steps, the most recent 20 steps, and the full history) affect code coverage. We take the history length of five steps as default, following the settings of GPTDroid~\cite{GPTDroid}. The results are reported in Table~\ref{tab:RQ3}.

\begin{table}[t]
\centering
\caption{Code coverage of GPTWeb under different history representations. The red and green arrows indicate whether the coverage in the current cell is lower or higher than that in the adjacent cell to its left.}
\resizebox{\linewidth}{!}{
\begin{tabular}{@{}c|c|ccc|ccc|ccc@{}}
\toprule
\textbf{}            & \textbf{}           & \multicolumn{3}{c|}{\textbf{Action History}}          & \multicolumn{3}{c|}{\textbf{State History}}           & \multicolumn{3}{c}{\textbf{Functionality History}}      \\ \midrule
\textbf{Application} & \textbf{No History} & \textbf{Len 5} & \textbf{Len 20} & \textbf{Unlim.} & \textbf{Len 5} & \textbf{Len 20} & \textbf{Unlim.} & \textbf{Len 5} & \textbf{Len 20} & \textbf{Unlim.} \\ \midrule
Dimeshift            & 32.85\%             & 21.41\%        & 38.43\%\green{$\uparrow$}         & 34.43\%\red{$\downarrow$}           & 30.13\%        & 27.96\%\red{$\downarrow$}         & 35.06\%\green{$\uparrow$}           & 36.72\%        & 42.11\%\green{$\uparrow$}         & 35.02\%\red{$\downarrow$}           \\
Pagekit              & 17.12\%             & 27.56\%        & 30.95\%\green{$\uparrow$}         & 26.07\%\red{$\downarrow$}           & 31.40\%        & 31.11\%\red{$\downarrow$}         & 20.84\%\red{$\downarrow$}           & 32.08\%        & 32.76\%\green{$\uparrow$}         & 27.27\%\red{$\downarrow$}           \\
Petclinic            & 20.00\%             & 26.00\%        & 54.00\%\green{$\uparrow$}         & 18.00\%\red{$\downarrow$}           & 38.00\%        & 38.00\%-         & 00.00\%\red{$\downarrow$}            & 54.00\%        & 56.00\%\green{$\uparrow$}         & 36.00\%\red{$\downarrow$}           \\
Phoenix              & 57.37\%             & 59.21\%        & 60.53\%\green{$\uparrow$}         & 65.53\%\green{$\uparrow$}           & 62.63\%        & 60.26\%\red{$\downarrow$}         & 56.58\%\red{$\downarrow$}           & 69.47\%        & 72.11\%\green{$\uparrow$}         & 65.00\%\red{$\downarrow$}           \\
Retroboard           & 45.85\%             & 63.27\%        & 61.64\%\red{$\downarrow$}         & 56.73\%\red{$\downarrow$}           & 56.96\%        & 55.67\%\red{$\downarrow$}         & 51.93\%\red{$\downarrow$}           & 62.81\%        & 62.22\%\red{$\downarrow$}         & 63.51\%\green{$\uparrow$}           \\
Splittypie           & 27.61\%             & 47.18\%        & 44.70\%\red{$\downarrow$}         & 40.94\%\red{$\downarrow$}           & 42.31\%        & 39.23\%\red{$\downarrow$}         & 36.32\%\red{$\downarrow$}           & 41.28\%        & 42.39\%\green{$\uparrow$}         & 41.20\%\red{$\downarrow$}     \\ \midrule
Average           & 33.47\%               & 40.77\%          & 48.38\%\green{$\uparrow$}                                        & 40.28\%\red{$\downarrow$}                                        & 43.57\%          & 42.85\%\red{$\downarrow$}                                        & 33.46\%\red{$\downarrow$}                                        & 49.39\%          & 51.27\%\green{$\uparrow$}                                        & 44.67\%\red{$\downarrow$}                                        \\
\bottomrule     
\end{tabular}}
\label{tab:RQ3}
\end{table}

By comparing different history representations (using the default ``Len 5'' columns), we observe that providing more historical information consistently improves testing effectiveness on average. For instance, functionality history increases the average absolute coverage by 15.92\% compared to no history. This improvement stems from the model’s enhanced ability to understand both the execution progress of the current functionality and its current page state when rich historical context is available, enabling effective action planning.

Under the setting of no history, the model makes purely single-step decisions. As a result, once it performs an incorrect action (e.g., clicking a non-interactable button), it tends to repeatedly execute the same erroneous action until the testing budget is exhausted. The same issue arises with action history. In contrast, when equipped with state history or functionality history, the model generates more diverse actions and gains the ability to explore alternative paths after observing repeated or unproductive actions, thereby achieving higher code coverage.

Under the settings of action history and functionality history, the average code coverage exhibits a trend of increasing and then decreasing as the length of history context grows. This pattern arises because longer histories provide the LLM with richer information for decision making, enabling a better understanding of the application under test.
However, when the retained context becomes excessively long, the rapidly expanding history can dominate the prompt, causing instructions that are relevant to the task to be overwhelmed. Moreover, overly long prompts can degrade the LLM’s reasoning capability~\cite{LostInTheMiddle}.
Although the action history setting shows instability due to retaining only action-level information, the functionality history setting demonstrates a clear rise-and-fall trend in coverage, highlighting the importance of selecting an appropriate length of history context.

Notably, when the history context length is unlimited, each request to the LLM retains all prior history, resulting in quadratic growth in prompt length. Consequently, the monetary cost under an unlimited history length is nearly 20× higher than that under a history length of 5. This result further demonstrates the practical importance of constraining history context length, not only for maintaining testing effectiveness but also for controlling computational and financial costs.

Under the setting of state history, code coverage generally decreases as the context length increases. This decline occurs because state history requires generating a natural-language description of the page after every action step, and these descriptions are often verbose and highly repetitive, resulting in substantially longer prompts than those under action history or functionality history.
We observe that when using state history with an unlimited context length, GPTWeb fails to achieve any code coverage (calculated in branch coverage) on Petclinic. Our analysis suggests that the LLM tends to repeat information that appears frequently in the prompt, and the redundant state descriptions exacerbate this tendency. As a result, GPTWeb repeatedly interacts with the same input field (i.e., entering and clearing text) instead of performing meaningful exploration.
These findings indicate that state history does not scale well to longer contexts, whereas providing concise, high-level functionality history constitutes a more effective and scalable way to supply historical knowledge to LLM-based AWGT approaches.

\textbf{Answer 3}: Providing sufficient contextual knowledge can substantially improve the testing effectiveness of LLM-based AWGT approaches. Such context should be presented in a concise yet high-level, functionality-rich form, rather than describing page states after every step. Selecting an appropriate history length is also critical. The context should be long enough to convey sufficient information, while avoiding excessive length that may degrade LLM effectiveness.

\section{RQ4 Failure-Revealing Capability}
\begin{figure}[t]
    \centering
    \includegraphics[width=\linewidth]{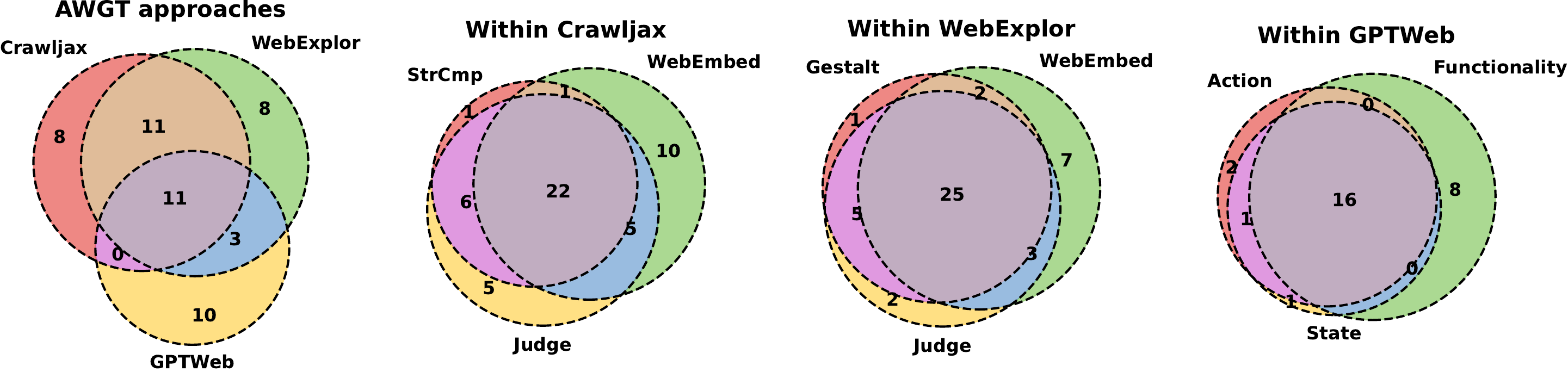}
    \caption{Venn diagrams comparing failures triggered by different AWGT approaches and failures triggered under different state abstraction techniques integrated within each AWGT approach.}
    \label{fig:RQ4Graph}
\end{figure}

In RQ4, we compare the unique failures revealed by different AWGT approaches under different integrated state abstraction techniques to assess their failure-revealing capabilities. The Venn diagrams are presented in Figure~\ref{fig:RQ4Graph}.

We follow a multi-step process to identify the unique failures triggered by each AWGT approach. First, we parse the logs generated by each approach to collect the complete set of failures. The logged failures are extracted from browser console errors. 
Second, we normalize failure messages by escaping newline characters and Unicode symbols, and by removing specific prefixes and suffixes from different AWGT approaches. 
Third, we remove URLs from failure messages using regular expressions, as URL parameters can interfere with deduplication. 
Fourth, based on manual inspection, we further eliminate variable parameters (e.g., inputs generated by AWGT approaches and numeric values in error messages) using regular expressions. 
Finally, we perform a string-based comparison to produce the final set of deduplicated unique failures.
Our processing script is publicly available in our repository~\cite{OpenSourceRepo}. Figure~\ref{fig:RQ4Graph} reports the total number of unique failures triggered by each AWGT approach across five runs on six applications.

As shown in the leftmost Venn diagram in Figure~\ref{fig:RQ4Graph}, WebExplor triggers the largest number of unique failures (33), whereas GPTWeb triggers the fewest (24). This result indicates that, although LLM-based approaches can achieve competitive code coverage, their failure-revealing capability remains limited by slow action execution.
However, GPTWeb exhibits low overlap with failures triggered by Crawljax and WebExplor, suggesting that LLM-based approaches can leverage LLM reasoning to probe deep application behaviors and uncover distinct classes of failures.

For Crawljax and WebExplor, we compare their default configurations with variants integrated with the most effective state abstraction techniques (Judge and WebEmbed). Although most failures overlap across different abstractions, each abstraction still reveals a set of exclusively triggered failures. Notably, WebEmbed accounts for the largest proportion of exclusively triggered failures, indicating that effective abstractions that construct fewer states can help AWGT approaches discover additional, previously unseen failures.

Comparing the total number of failures with average code coverage reveals no strong correlation between these two metrics. For Crawljax, integrating StrCmp, WebEmbed, and Judge yields average coverages of 49.12\%, 45.55\%, and 54.18\%, while triggering 30, 38, and 38 unique failures, respectively. Similarly, for WebExplor, integrating Gestalt, WebEmbed, and Judge achieves average coverages of 52.45\%, 53.45\%, and 51.78\%, while revealing 33, 37, and 35 unique failures. These results suggest that higher coverage does not necessarily translate into stronger failure-revealing capability.

The rightmost Venn diagram shows that GPTWeb equipped with functionality history uncovers the largest number of unique failures and produces the highest number of exclusively triggered failures. In contrast, failures revealed under action history and state history are almost entirely subsumed by those revealed under functionality history. This finding demonstrates that LLM-based AWGT approaches strongly benefit from semantically rich historical context, enabling effective exploration and deep failure discovery.

\textbf{Answer 4}: Code coverage shows no strong correlation with failure-revealing capability. Different categories of AWGT approaches expose complementary failure sets. RL-based approaches detect the largest number of unique failures, while LLM-based approaches uncover the most exclusively triggered failures. Meanwhile, state abstraction techniques substantially influence failure discovery, with techniques that balance abstraction effectiveness and compact state representations (e.g., WebEmbed) yielding the greatest benefits for model-based and RL-based approaches. In contrast, LLM-based AWGT approaches benefit most from semantically rich historical context, where functionality history is demonstrated to be the most effective.

\section{Threats to Validity}
\textbf{Internal Validity.}
First, the parameter settings and configurations of the approaches and techniques under study may affect the evaluation results. To mitigate this threat, we adopt the recommended settings for each approach and technique as reported in prior work~\cite{WebExplor,crawljax,fraggen,GPTDroid,webembed,Judge}, and configure all approaches with identical login scripts and the same time budget.
Second, faults may exist in the evaluation framework of our study. We mitigate this threat by using the open-source implementations of the studied approaches and techniques and conducting thorough code reviews.
Third, randomness may threaten the validity of our evaluation. We mitigate this threat by repeating each experiment five times and reporting the average results, following standard practice in prior studies~\cite{WebRLED,Judge,WebQT,WebExplor}.

\textbf{External Validity.}
The selection of web application subjects may introduce bias. To mitigate this threat, we adopt representative open-source web applications that have been widely used in prior studies~\cite{DIG,WebExplor,WebQT,Judge}. Our evaluation results indicate that these subjects exhibit sufficient diversity and complexity to reveal meaningful differences and trends among the approaches under study.

\section{Related Work}

\subsection{Automated Web GUI Testing}
Automated web GUI testing has been widely studied by various approaches.
Random-based approaches include Monkey~\cite{monkey}, White et al.~\cite{visualRandom}.
Model-based approaches include Crawljax~\cite{crawljax}, ATUSA~\cite{ATUSA}, TESTAR~\cite{testar}, GUITAR~\cite{guitar}, FeedEx~\cite{FeedEx}.
RL-based approaches include WebExplor~\cite{WebExplor}, WebQT~\cite{WebQT}, QExplor~\cite{QExplore}, UniRLTest~\cite{UniRLtest}, PIRLTest~\cite{PIRLTest}, and WebRLED~\cite{WebRLED}.
These approaches do not require a pre-constructed model or the source code of the AUT, and can be directly applied to test the AUT in the form of black-box testing.
In this study, we select representative AWGT approaches to study the differences among AWGT approaches according to their code coverage.

\subsection{LLMs for GUI Testing}
Many recent approaches apply LLMs for automated GUI testing by exploiting their strong reasoning capabilities. QTypist~\cite{QTypist}, InputBlaster~\cite{InputBlaster}, and VETL~\cite{VETL} include LLMs to generate meaningful text inputs for RL-based GUI testing approaches, thereby improving test effectiveness.
Guardian~\cite{guardian}, AutoE2E~\cite{AutoE2E}, and NaviQAte~\cite{naviqate} focus on functional testing by automatically inferring key application functionalities and using LLMs to exercise them.
GPTDroid~\cite{GPTDroid}, DroidAgent~\cite{DroidAgent}, MobileGPT~\cite{mobileGPT}, and Trident~\cite{DroidAgent} rely on LLMs to make every action decision during testing.
LLMDroid~\cite{llmdroid} and Temac~\cite{temac} integrate LLMs to assist traditional GUI testing approaches in further improving code coverage.
However, approaches designed for mobile testing often struggle when applied to web applications, due to the complexity and dynamic behavior of modern web applications~\cite{VETL,WebExplor,WebQT}.
In this study, we migrate GPTDroid to the web scenario, and equip it with four different history representation strategies to study their impact on code coverage.

\subsection{State Abstraction}
State abstraction is essential and widely applied in AWGT approaches, enabling the detection of already-tested pages and thereby avoiding redundant testing.
NDStudy~\cite{NDStudy} first studied the effect of threshold-based state abstraction techniques on guiding AWGT approaches to build abstract models of web applications. The studied techniques are either based on HTMLs as input~\cite{RTED,Levenshtein,SimHash}, or based on screenshots as input~\cite{PDiff,PHash,SIFT}. 
Several variants of threshold-based techniques adopt multi-modal techniques~\cite{fraggen} and heuristic rules~\cite{WebExplor} to improve effectiveness.
Recent techniques adopt machine learning and deep learning techniques for state abstraction~\cite{TK,webembed,Judge}.
In this study, we choose six different state abstraction techniques with different AWGT techniques to study the interaction between state abstraction and exploration strategy.

\section{Conclusion}
In this article, we present an empirical study of automated web GUI testing (AWGT), examining the joint affect of exploration strategies and state abstractions on testing effectiveness. 
Our results show that no single AWGT approach consistently outperforms others. 
Instead, different approaches exhibit complementary strengths in coverage and failure discovery. 
We find that fine-grained state abstractions benefit model-based strategies, while compact abstractions better support RL-based strategies. 
For LLM-based AWGT approaches, concise and functionality-level history representations are most effective, whereas verbose state-level histories degrade effectiveness. 
We further observe that code coverage is not strongly correlated with failure-revealing ability, highlighting the need to evaluate both dimensions. 
Overall, our findings provide actionable guidance for designing, configuring, and evaluating AWGT approaches, paving the way toward more effective automated web GUI testing.



\bibliographystyle{ACM-Reference-Format}
\bibliography{sample-base}

\appendix

\end{document}